\documentclass[twocolumn,           
               showpacs,            
               preprintnumbers,     
               aps,                 
               prl,                 
               a4paper,             
               superscriptaddress,  
               nofootinbib,         
               tightenlines,        
               floats,floatfix      
               ]{revtex4}

\usepackage{graphicx}
\usepackage{dcolumn}
\usepackage{bm}
\usepackage{latexsym}
\usepackage{amsmath,amssymb}        
\usepackage[draft=false]{hyperref}

\usepackage[latin1]{inputenc}
\usepackage{latexsym}
\usepackage{amsmath}
\usepackage{epsfig}
\usepackage{ifpdf}

\begin{document}
\title{Scalar Field Dark Matter Quantum Effects as Dark Energy}
\author{Tonatiuh Matos\footnote{Part of the Instituto Avanzado de Cosmolog\'ia (IAC) collaboration http://www.iac.edu.mx/}}
\email{tmatos@fis.cinvestav.mx}
\affiliation{Departamento de F\'isica, Centro de Investigaci\'on y de
  Estudios Avanzados del IPN, A.P. 14-740, 07000 M\'exico D.F.,
  M\'exico.}

\begin{abstract}
Using very simple arguments we show that the quantum effects of an ultralight particle as the Scalar Field Dark Matter $m_{SFDM}\sim10^{-22}$eV cannot be neglected at classical scales. We show that the effective density of this effect is constant in space and for such a mass, it is of the order of magnitude of the critical mass of the universe. Thus, we can interpret the effective density of this quantum effects as the cosmological constant. 
\end{abstract}

\date{\today}


\maketitle

Doubtless one of the greatest misteries science is now facing on is the nature of the dark energy \cite{Xia:2009ys}\cite{Perlmutter:1999rr}. It represents three fourths of the whole matter of the universe and its physical nature seems to contradict all the known physics. The best accepted candidate is the cosmological constant $\Lambda$, introduced by Albert Einstein in the 1920's \cite{rewiew}. People use to interpret $\Lambda$ as fluctuations of the vacuum energy, nevertheless this interpretation does not coincide with value obtained using any quantum field theory, where the vacuum energy is zero or 128 orders of magnitude bigger than the observed one in nature \cite{SantosCorchero:2009zz}. On the other hand, dark matter continues being a puzzle of the cosmology as well. The best candidates, supersymmetric particles, seem to evade the detectors and people is starting to think in alternatives. However, the Scalar Field Dark Matter (SFDM) \cite{Matos:1998vk} remains as one of the best candidates, because: 1.- The SFDM model has only one free parameter, the scalar field mass $m_{SFDM}$ \cite{Matos:2000ss}, 2.- The ultralight scalar field mass ($m_{SFDM} \sim 10^{-22}$eV) fits: the observed amount of substructure \cite{Matos:2000ss}, the critical mass of galaxies \cite{jaeweon},\cite{Alcubierre:2002et}, the rotation curves of galaxies \cite{Boehmer:2007um}, the central density profile of LSB galaxies \cite{Bernal:2005RMAA}, the evolution of the cosmological densities \cite{Matos:2008ag}, etc. 3.- SFDM forms galaxies earlier than the Cold Dark Matter (CDM) model, because they condensate in Bose-Einstein Condensates at a critical temperature $T_c >> $TeV \cite{Matos:2008ag}. So, if SFDM is right, we have to see big galaxies at big redshifts. 

Thus, SFDM particles have a mass of $m_{SFDM}\sim10^{-22}$eV, with this mass we expect that the quantum mechanical effects are important at the scale of $\sim1/m_{SFDM}\sim 1$pc. In this letter we show that if the SFDM are the dark matter, quantum effects of the ultralight particles can explain a constant density on the background of the universe and can be interpreted as a cosmological constant. 

In order to do that, suppose that we observe a big region of the universe, as big as our horizon, in a time-scale much smaller than the Hubble-time. Halos of galaxies can be seen as perturbations of the Einstein-Klein-Gordon (EKG) equations $R_{\mu\nu}=\kappa\Phi_\mu\Phi_\nu-2V(\Phi)$. We focus on  a perturbations $\delta\Phi$ of the scalar field $\Phi=\Phi_0+\delta\Phi$, where $\Phi_0=\Phi_0(t)$ determines the cosmological evolution of $\Phi$. In the Newtonian gauge $ds^2 =  -(1+2\psi)dt^2+a^2[(1-2\phi)\delta_{ij}dx^idx^j]$, the  EKG equations reduce to the Schr\"odinger-Poisson system \cite{Hu:2000ke}
\begin{subequations}
\begin{eqnarray}
 && i\,\hbar \dot\Psi+ +\frac{\hbar^2}{2\,m}\nabla^2\Psi-m\,c^2\Psi\,\phi_N =0 \label{eq:KK_Psia1a}
\\
&&\nabla^2\phi_N=4\pi G\,\frac{m^2\,c^2}{\hbar^2}\,\Psi\Psi^*
\label{eq:KK_Psia1b}
\end{eqnarray}\label{eq:KK_Psia1}
\end{subequations}
where $\delta\Phi=\frac{1}{\sqrt{2}}\Psi\,\exp(-i\,m\,c/\hbar\,t)+c.c.$, $\phi=\phi_N+\phi_1\,\exp(-2\,i\,m\,c/\hbar\,t)+c.c.$
Remember that we are neglecting the time evolution of the universe for the moment ($\dot a=0$). Now we write the Schr\"odinger equation in its hidrodynamial version, 
\begin{subequations}
 \begin{eqnarray}
 \frac{\partial\rho}{\partial t}+\nabla \cdotp\left(\rho \textbf{v} \right) &=&0\\
\frac{\partial(\rho\textbf{v} )}{\partial t}+\nabla \cdotp\left(\rho \textbf{v}\textbf{v} \right)&=&\rho\,F_Q-\frac{\rho}{m}\nabla \phi_N\\
\nabla^2\phi_N&=&4\pi G\,\frac{m^2\,c^2}{\hbar^2}\,\rho
\end{eqnarray}\label{eq:hydro}
\end{subequations}
Equations \eqref{eq:hydro} are equivalent to the Schr\"odinger-Poisson equations if the wave function $\Psi$ is decomposed as $\Psi=\sqrt{\rho}\,\exp(iS)$, being $\textbf{v} =\hbar/m\nabla S$. The main difference of equations \eqref{eq:hydro} and the hydrodynamical equations is the quantum force $F_Q=-\nabla U_Q$, which is derivable from the quantum potential
\begin{equation}
 U_Q=-\frac{\hslash^2}{2\,m^2}\frac{1}{\sqrt{\rho}}\left( \nabla^2\sqrt{\rho}\right)
 \label{eq:U_Q_1} 
\end{equation}
$U_Q$ is interpreted as the quantum mechanical contribution to the hydrodynamical equations. If we drop out the quantum force $F_Q$ in \eqref{eq:hydro}, these equations become the classical ones for a hydrodynamical system. Observe that the system \eqref{eq:hydro} without the quantum contribution is equivalent to the CDM model, the main difference is that in CDM the dark matter is modeled as a fluid, but it is shown in \cite{widrow} that the result at the cosmological level is the same as in the case of the scalar field. Thus the main difference between \eqref{eq:hydro} and CDM is the quantum term $F_Q$ in \ref{eq:hydro}. In what follows we only focus on this term.
Of course, the universe is quantum mechanical but the contribution of $F_Q$ to the hydrodynamical equations is very small due to the factor $\frac{\hslash^2}{2\,m^2}$, which make this force very small for a mass $m$ of a normal classical object. But for a SFDM field of mass $m_{SFDM}\sim 10^{-22}$eV this term can not be neglected anymore. 

Of course in the very early universe, the SFDM field was homogeneous and isotropic, but later, if it is the dark matter in the universe, it is not longer uniformly distributed in the universe, the dark matter forms the halos in clusters of galaxies, galaxies, etc. We can estimate the density or the square root of the density in the following way. 

The distribution of the dark matter can be modeled as concentrations of matter. In order to solve \eqref{eq:hydro}, we can make a comparative analysis if we observe what happens with concentrations as Dirac delta functions distributions $\sqrt{\rho}\sim\delta(r-r_0)$. If we accept the cosmological principle, the dark matter clumps are homogeneously distributed in the universe, one can suppose that if there exists a galaxies in $r_0$, it should exist another galaxy in more or less $-r_0$, thus  we write that $\sqrt{\rho}\sim\delta(r-r_0)+\delta(r+r_0)=2\,|r_0|\delta(r^2-r^2_0)$. It is well known that the delta function can be written as $\delta(a\,r^ 2-1)=\underset{l\rightarrow\infty}\lim n\,J_n\,(a\,n\,r^2)$, where $J_l$ are the Bessel functions. 

Of course, the halos of galaxies have a smooth distribution more than a delta function like one. Nevertheless, we can use the last idea to expand the density in Fourier series. What we learn is that it is convenient to expand the density distribution of the haloes of galaxies as
\begin{equation}
 \sqrt{\rho}=\sum_{l\,m\,n}U_{l\,m\,n}\frac{J_{n}(k^2\,r^2)}{\sqrt{r}}Y^l_{m}(\theta,\,\varphi)\label{eq:rho}
\end{equation}
where $k$ is the wave number of the scale of fluctuations of the dark matter haloes with units of $L^{-1}$ and $Y^l_m$ are the spherical harmonic functions. Of course, any function can be expanded as equation  \eqref{eq:rho}, because the Bessel and the spherical harmonic functions are basis of the functions space. Expansion \eqref{eq:rho} formally represents the distribution of the galaxies haloes in the horizon size, fortunately for our goal we do not need to calculate it exactly, it is enough to know its expansion only. Now we are able to calculate the quantum potential $U_Q$, using expansion \eqref{eq:rho} and after some simple algebra we obtain
\begin{equation}
 U_Q=-\frac{\hslash^2}{2\,m^2}\frac{1}{\sqrt{\rho}}\left( \nabla^2\sqrt{\rho}\right) = \frac{2\,\hslash^2}{m^2}k^4\,r^2 + O\left(\frac{1}{r^2}\right)\label{eq:U_Q}
\end{equation}
$U_Q$ is the potential that effectively feels an observer from the galaxies due to the quantum effects of the SFDM. An effective density can be obtained with the Poisson equation $\nabla^2\,U_Q = 4\pi\,G\,\rho_{eff}$ using  \eqref{eq:U_Q} as a source of the density. We get
\begin{equation}
 \rho_{eff}=12\frac{k^4\,\hslash^2}{\pi\,G}\frac{1}{m_{SFDM}^2}+ O\left(\frac{1}{r^4}\right)\label{eq:rhoeff}
\end{equation}
 
The first result we observe here is that the first term of the effective density is constant, it does not depend on the position. The other terms which go like $O(1/r^4)$ are important only locally, thus, we can interpret them as the contribution of the nearby structure. This is a remarkable result, because we see that quantum effects cannot be neglected for masses of the order of the SFDM mass and for large scales, its effective density is constant. The second result is that this effect appears just when the structure begins to form, like as that we expect for the cosmological constant. Nevertheless, this effective density does not have to be of the size of something well-know. 

We set the observed values of the different quantities in this formulae. Observe that for big structures, $k=2\pi/L$ is small, thus, the leader term in expression \eqref{eq:rhoeff} corresponds to the smallest structure. The smallest structure we observe in the galaxies are the dwarf and satelites galaxies, with a size of $\sim10$kpc, thus $k\sim1.5\cdotp10^{-37}\text{GeV}$ (in unites where $\hbar=c=1$), $m_{SFDM}\sim10^{-22}$eV, we obtain that
\begin{equation}
 \rho_{eff}\sim4\left(1.5\cdotp10^{-37}\text{GeV}\right)^4 \frac{m_{Pl}^2}{m_{SFDM}^2}\sim 10^{-47}\text{GeV}^4
\end{equation}
This is again a remarkable surprise, the effective density due to the quantum effects of a particle like the SFDM is of the order of the critical density of the universe $\rho_{crit.}=8.099\,h^2\,10^{-47}$GeV$^4$, that means, it is possible that the constant energy we feel in the universe is due quantum effects of small particles of dark matter.

In this letter we have shown that, using very simple calculations,  a particle of a small mass has quantum effects at classical scales. This is not a surprise, because we know that the quantum effects of a particle are of the order of the de Compton length, inverse to the mass of the particle $\sim1/m$. The first remarkable result is that with very simple arguments but using the hydrodynamical version of the Schr\"odinger equation, we can see that the effective density of the quantum effects is constant. The great surprise is that the amount of this density is just of the order of magnitude of the critical density, which coincide almost with the amount of dark energy in the universe. 

Observe in (\ref{eq:U_Q_1}) that there is no contribution of these quantum effects as long as the quantum density $\rho$ is homogeneous. But, it is important when the structure of the universe begins to form. This is a very elegant way to solve the coincidence problem.

Of course, the fact that we do not see the supersymmetric particles in the detectors is not an argument to discard them as dark matter. The SFDM is also a very good candidate to be the dark matter, fortunately without the problems of the WIMPs \cite{Batell:2009vb}, like the cuspy central densities in galaxies, the excess of substructure, etc. (see for example \cite{Matos:2000ss},\cite{Bernal:2005RMAA},\cite{Matos:2008ag}). However, if the SFDM remains as candidate of the dark matter of the universe, we have to take into account the quantum effects of this particle in the background, they cannot be neglected. 

This work was partially supported
by CONACyT M\'exico under grants 49865-F and I0101/131/07
C-234/07 of the Instituto Avanzado de Cosmologia (IAC) collaboration
(http://www.iac.edu.mx/).


\begin{thebibliography}{99}

%
\bibitem{Xia:2009ys}
  J.~Q.~Xia and M.~Viel,
  JCAP {\bf 0904}, 002 (2009)
  [arXiv:0901.0605 [astro-ph.CO]].
\bibitem{Perlmutter:1999rr}
  S.~Perlmutter,
in  {\it Proc. of the 19th Intl. Symp. on Photon and Lepton Interactions at High Energy LP99 } ed. J.A. Jaros and M.E. Peskin,
%
\bibitem{rewiew} Ruth Durrer. The Cosmic Microwave Background. Cambridge University Press (2008).
%
\bibitem{SantosCorchero:2009zz}
  E.~Santos Corchero,
  AIP Conf.\ Proc.\  {\bf 1122}, 229 (2009).
%
\bibitem{Matos:1998vk}
  T.~Matos and F.~S.~Guzman,
  Class.\ Quant.\ Grav.\  {\bf 17}, L9 (2000)
  [arXiv:gr-qc/9810028].
  %
\bibitem{Matos:2000ss}
  T.~Matos and L.~A.~Urena-Lopez,
  Phys.\ Rev.\  D {\bf 63}, 063506 (2001)
  [arXiv:astro-ph/0006024].
  %
\bibitem{jaeweon} Jae-Weon Lee. Phys.\ Rev.\ D {\bf 53}, 2236 (1996)
  %
\bibitem{Alcubierre:2002et}
  M.~Alcubierre, F.~S.~Guzman, T.~Matos, D.~Nunez, L.~A.~Urena-Lopez and P.~Wiederhold,
  arXiv:astro-ph/0204307.
  %
\bibitem{Boehmer:2007um}
  C.~G.~Boehmer and T.~Harko,
  JCAP {\bf 0706} (2007) 025
  [arXiv:0705.4158 [astro-ph]].
%
\bibitem{Bernal:2005RMAA}
A. Bernal, T. Matos and D. N\'u\~nez. 
Rev. Mex. A.A. 44, (2008), 149-160. E-print: astro-ph/0303455.
%
\bibitem{Matos:2008ag}
  T.~Matos, J.~A.~Vazquez and J.~Magana,
MNRAS, {\bf 389}, 13957 (2009)
  arXiv:0806.0683 [astro-ph].
%
\bibitem{Hu:2000ke}
  W.~Hu, R.~Barkana and A.~Gruzinov,
  Phys.\ Rev.\ Lett.\  {\bf 85}, 1158 (2000)
  [arXiv:astro-ph/0003365].
  %
  \bibitem{widrow} L. Widrow and N. Kaiser. ApJ {\bf 416},  L71 (1993) 
%
\bibitem{Batell:2009vb}
  B.~Batell, M.~Pospelov and A.~Ritz,
  Phys.\ Rev.\  D {\bf 79}, 115019 (2009)
  [arXiv:0903.3396 [hep-ph]].
%
\end{thebibliography}
\end{document}